\documentclass{llncs}

\usepackage{url}

\newcommand{\keywords}[1]{\par\addvspace\baselineskip
\noindent\keywordname\enspace\ignorespaces#1}
\newcommand{\ie}{\emph{i.e.}}
\newcommand{\eg}{\emph{e.g.}}
\begin{document}
\title{Design-on-demand or how to create a~target-oriented social web-site}

\date{March 27, 2009 (v. 0.6)}

\author{Jaros{\l}aw Adam Miszczak\inst{1} \and Izabela Sobota-Miszczak\inst{2}}
\institute{Institute of Theoretical and Applied Informatics,\\ Polish Academy
of Sciences,\\ Ba{\l}tycka 5, 44-100 Gliwice, Poland\\
\email{miszczak@iitis.pl}
\and 
ASLAN Company,\\ Kasprzaka 1/17, 44-121 Gliwice, Poland}
\authorrunning{J.A. Miszczak, I. Sobota-Miszczak} 

\maketitle

\begin{abstract}
We describe an informal methodology for developing on-line applications, which
is, to some extent, complementary to the Web 2.0 aspects of web development. The
presented methodology is suitable for developing low-cost and non-cost web sites
targeted at medium-sized communities. We present basic building blocks used in
the described strategy. To achieve a better understanding of the discussed
concepts we comment on their application during the realization of two web
projects. We focus on the role of community-driven development, which is crucial
for projects of the discussed type.
\keywords{web communities, design methodology}
\end{abstract}

\section{Introduction}
As Internet constantly becomes increasingly important medium for providing
content and sustaining relations between people, more attention is focused on
the web-development methodology~\cite{ginige01webengineering,lee04implementing}.
The methods presented in the literature, however, are rarely used in the
real-world projects~\cite{jeary09evaluation,lang05hypermedia,glass97revisiting}.
There are also some points to support the claim that the web development process
is in many areas not as different from the~traditional development process as
suggested in the literature~\cite{lang05hypermedia}. Clearly, there is a gap
between the new methods proposed in the academic community and the pragmatic way
of thinking required in order to realize any web-project in a business
environment or any other real-world enterprise~\cite{glass97revisiting}.

The problem of web-design methodology has been extensively studied in the
literature (see \eg~\cite{jeary09evaluation,lang05hypermedia} and references
therein). Among the proposed methodologies there are some which aim to be user
centric~(\eg~\cite{troyer01audience}) rather than data-centric. The main
attention in the proposed methodologies, however, is focused on the
software development process and, in most cases, this process is
considered to be finalized after the publication of software. We are not aware
of the methodology aimed at long-term development of web-communities. This area,
however, has recently gained some attention in the context of social web
applications~\cite{bell09building}.

Moreover, most of the methods described in the literature are focused on the
complex business projects. Researchers seem to ignore the large audience of
amateur web-developers and non-commercial communities, which become more and
more important for the landscape of the Internet. One should note that this
trend cannot be fully described in such terms as \emph{Web 2.0}~\cite{web20} and
\emph{user-generated content}~\cite{wunsch07participative}. The reason for this
is that during the recent years it has become more popular not only to use
provided services to create content, but also to create customized,
topic-oriented web sites using web content management systems
(CMS)~\cite{mehta-cms} or wikis~\cite{wiki}.

The main aim of this paper is to provide an overview of the methodology which
can be helpful for anyone planning to develop and maintain a website targeted at
a medium-sized community. Our main goal is to identify basic ingredients which
we find quintessential for the process of creating such websites and web
communities. We do not aim to provide a formal methodology for creating a
full-featured web-project. On the contrary, we rather aim at bringing to your
attention the most important aspects of such projects, with the special focus on
the role of a community. Thus, our paper can be viewed as a mini-howto, rather
than the comprehensive description of the methodology~\cite{kawasaki04art}.

We focus our attention on medium-sized web communities and we present two case
studies of web projects developed in such groups. The first one is the case of a
portal for quantum information scientific community and the second one aims at
providing support for independent travellers in Turkey. These two cases
illustrate our theory, as tourism or information technology are examples of the
large community areas of interest, while independent travels in Turkey or
quantum computing seem to be aimed at medium-sized ones.

This paper is organized as follows. In Section \ref{sec:main} we introduce the
basic elements of the successful process which aims at building a community
based web-site. We also briefly comment on the relation of introduced concepts
with the existing methodologies concentrated on developing web-sites. Section
\ref{sec:build} shortly describes the difference between the process of creating
a social web-site using the introduced components and the sites built by using
other approaches. In Section \ref{sec:examples} we describe two real-world
examples of successful web communities created by using the presented
methodology. Finally Section \ref{sec:final} gives some concluding remarks and
comments on the relations between the described methodology and some existing
ones.

\section{Components}\label{sec:main}
We start with the list of components or requirements which are crucial for the 
presented methodology. The detailed discussion of below listed ingredients can
be found in further subsections.

\newcounter{component}
\begin{list}{\textbf{(C\arabic{component})}}{\usecounter{component}}

  \item \textbf{Subject -- focused and expandable}
  \label{com:subject}\par 
  The basic issue in creating a ''design-on-demand'' website is the selection of
  the main topic for the site. As the aim is a medium-sized community the main
  topic should be wide enough to attract a significant number of visitors per
  day (\ie\ several thousand visitors a day). At the same time it should be
  focussed and limited by clearly defined borders.

  \item \textbf{People -- team and supporters}
  \label{com:team}\par
  The most important issue is to establish a small yet creative and
  target-driven team of website developers and key users. The developers are the
  core value of the site since they are the people who create the general layout
  of the site and its initial content. Main users or supporters are considered
  to be crucial at the second stage of site development, because they provide
  the developers with necessary feedback and support as well as criticism of the
  original ideas implemented in the site-development stage.

  \item \textbf{Promotion -- low-cost and organic}
  \label{com:promotion}\par
  Promotion and positioning are frequently associated with profit-generating
  activities. However, even in the case of non-profit web site, its promotion is
  a key factor that allows the site to gather a community of supporters and,
  thus, grow steadily.

  \item \textbf{Technological background -- flexible and extensible engine} 
  \label{com:flex}\par
  The choice of a tool used during the development is also very important part
  of the process. In fact this part is crucial, as it determines the ability of
  the project to evolve. As we aim to address the problem of building low-cost
  sites, the most natural choice for the development tool is a content
  management system or a wiki available under the terms of one of the free
  software licenses.

  \item \textbf{Long-term involvement} 
  \label{com:long}\par 
  The methodology described in this paper was tested by the authors during
  the development of non-commercial web-sites. In such cases one should expect
  rather slow development of an undertaken project. Thus, it is crucial to be
  able to attract and retain the attention of contributors and project
  initiators as one should expect rather long-term involvement.

  \item \textbf{Openness for the community expressed in the initial design}
  \label{com:open}\par
  The openness element is the one explaining the ''design-on-demand''
  designation. For the project to attract new users and editors, it has to be
  open for a feedback from the community. People managing the project should be
  able to implement new features requested by the users. In this way the project
  becomes a collaborative effort and its shape (web layout, provided
  functionality) is adjusted for the needs of a community. The openness
  motivates the need for a flexible engine. It also requires long-term
  involvement.

\end{list}

\subsection{Subject (C\ref{com:subject})}
As we have mentioned before, the choice of the appropriate subject is one factor
that determines the success or the failure of the project. Medium-sized
community is usually built around one common area of interest, which is too
focused to attract the attention of millions of Internet users. On the other
hand, it should be targeted at limited and well-defined groups of people, which
are large enough to sustain constant development and improvement of the project.

The first question one should consider before committing to the web project is
the one about the target audience. It is important to understand what is the
expected size of the community and what are their expectations.

To find out the most important expectations of our target audience it is
necessary first to check if similar websites already exist on the web and, if
so, what is missing from their content. If there are no websites on the given
subject one should research the expectations of the potential users and visitors
by browsing forums and blogs in the search of frequently asked questions which
tend not to receive significant attention.

Blogs and forums, however, present the very wide scope of knowledge in an
incoherent way. One can be faced with hundreds of answers to their question
presented randomly and chaotically in the case of a typical forum. On the other
hand blogs tend to be very subjective and biased.

In contrast, the ''design-on-demand'' website is a platform for presenting
knowledge and expertise in an organised and systematic way, while remaining open
for suggestions and enquiries from non-experts. Thus, typical
''design-on-demand'' site will differ from Web 2.0 sites, as it is not
completely governed by the users. This allows to sustain the higher quality of
presented content.

\subsection{People (C\ref{com:team})}
The second component one should take into consideration is the team of
developers. The optimal number of developers should not exceed 3-4 persons. This
results from the basic aspects of social group dynamics. When the number of
developers is greater the conflicting ideas about the general design and content
arise, and these conflicts, in spite of being inspiring, prevent the website
from being developed efficiently and seamlessly.

We suggest that the project developers should rather have different areas of
expertise, \ie\ technology, editorial and content-development
issues~\cite{kawasaki04art}. This allows for the better organization of
responsibilities among the core team. Otherwise the conflicting perspectives of
developers will more likely be destructive and discouraging for the whole
initiative.

One can distinguish four groups of people related to the typical
''desing-on-demand'' project.
\begin{enumerate}
  \item \emph{Developers} -- the core team (2-4 people) responsible for the
  system administration and content editing.
  \item \emph{Main users} -- the followers of the website responsible for
  contributing significant portions of content and suggesting improvements.
  \item \emph{Secondary users} -- people sporadically visiting the web site, but
  nevertheless providing some feedback and suggesting new features.
  \item \emph{Consumers} -- people visiting the site who never provide any
  feedback.
\end{enumerate}

We should stress here the importance of community for the ''design-on-demand''
project. As pointed above \emph{main users} and \emph{secondary users} groups
represent the people who promote the site externally, provide critical feedback
to the developers and suggest the possible new openings and issues that the site
should address. These two groups are responsible for the evolution of the web
project and most of the new features are suggested by the people from these
groups.

The core of the community is formed by the main and the secondary users. The
most important difference between these groups is the wiliness to contribute the
content. It is expected that the main users have the knowledge required to
participate actively in the discussions concerning the subject of the web site.
They are also responsible for content contributions. On the other hand the
secondary users are frequently people willing to learn about a new subject. As
such they can help to clarify some topics, by \eg\ asking questions,
participating in mailing lists. Their input is valuable for it helps to sustain
high-quality of the provided content.

Most of the project visitors are in the \emph{consumers} group. They are not
tightly connected with the community, but nevertheless they are interested in
the web-site subject. This group has some influence on the project as the
developers can observe, by the means of web analytics software, the search
trends and modify the provided content accordingly.

\subsection{Promotion (C\ref{com:promotion})}
In order to build a sustainable community of supporters it is crucial to attract
some critical number of visitors to the newly-created site. The methods of
promotion are similar to the ones used by commercial projects. However, as it is
assumed that the site is developed by a passionate, not by a business person,
they should be no-cost or low-cost methods.

The basic rules of organic positioning are clearly stated: use keywords,
register the site with major catalogues, remember about the site
map~\cite{seo-art}. However, there are other methods that the creators of
''design-on-demand'' site can use to attract the visitors and, what's more,
establish their credibility within the community of internet users as the
experts in a given subject.

The first method is to create the profile related to the developed site in
well-known community portals, such as Facebook or MySpace, as well as use the
forms of social communication offered by the web applications such as Twitter.
The common idea of providing RSS feed concerning the latest developments on the
site, which is also of a great value, seems to be the prototype of these
applications. 

The second approach, not so obvious as the first one and more time- and
effort-consuming, is to find some platforms of dialogue with potential visitors
and supporters. These platforms could be: forums, mailing lists and community
portals devoted to a subject broader than the focus point of our site. 

Building the image of an expert in a given area by the means of available
internet platforms calls for constant involvement and in-depth knowledge of the
subject, yet these requirements are not satisfactory. Interpersonal skills,
sympathetic approach and, above all, experience in dealing with internet
communities are the desired skills.

\subsection{Engine (C\ref{com:flex})}
We have already pointed out that content management systems and wikis are the
most natural tools which can be employed in order to develop a design-on-demand
web site. One should note that our main focus are topic-oriented projects and
for such the content is the most important element. Thus, the initiators and the
users of the project should focus on the ability to extend provided content. As
it is expected that the project should evolve, it is crucial to choose a
flexible and extensible one.

Before staring a successful ''design-on-demand'' project it is important to
understand the limitations of the chosen engine~\cite{mehta-cms,wiki}. Two
examples presented in Section~\ref{sec:examples} are built with the use of the
content management system. The important feature of CMS is its ability to
distinguish different groups of users by specifying privileges. Such mechanisms
are not always present in wikis.

Simultaneously one should not focus on the visual aspect of the selected engine.
For the popular engines it is easier to find ready-to-use layouts prepared by
other users. The larger user-base for the selected engine also facilitates
troubleshooting.

The content and the possibilities of its development are far more critical than
sophisticated options and graphical potential of the given engine. There are
some elements however, that allow the interaction between the developers and
users, which should be included in any project. First of all, the possibility of
adding comments and contacting developers is crucial for the purpose of
gathering users' suggestions. Secondly, elements such as forum or other form of
submitting content, allow new users to create small segments of content. This
provides the means for discovering users' interests and discussing suggestions.

\subsection{Involvement (C\ref{com:long})}
The ''design-on-demand'' philosophy of website creation requires the long-term
involvement of the developers team. Contrary to the short-term-involvement of
website designers whose goal is to launch the site and move on to the other
project, the ''design-on-demand'' project calls for long-term involvement of its
creators. It results, naturally, from the initial assumption: the site will grow
and adapt to the requirements of its users. The users ask questions and the
developers respond to them. Thus, the web site development becomes a continuous
dialogue between the site's users and designers.

The ''design-on-demand'' method has one significant advantage over the business
applications: the lack of deadlines. While business website developers are
forced to work under the pressure of time, the ''design-on-demand'' designers
have virtually unlimited time span. This fact, however, can result in decreased
motivation over the longer period of time. It is generally difficult to focus
one's attention to one project over an extended time-horizon. This requires some
motivating factors, such as the feeling of commitment, satisfaction and being
useful for the wider community. 

Another aspect that distinguishes the typical ''design-on-demand'' project from
a commercial one is the involvement of the developers. For this group the site
is usually a side activity or strictly a hobby. This fact results in several
consequences. Firstly, the time they spend on site development is strictly
limited, but productively used. Secondly, their emotional involvement in the
site is typically high, otherwise they would not get involved in the project.
Thirdly, it may be a challenging task to keep them interested in the site
development over prolonged periods of time.

The selection of the subject that attracts a medium-sized community is an
advantage here, having taken into account the limitations of site's developers.
The site will never attract millions of users and thus the full control of
site's content is possible to be maintained.

\subsection{Openness (C\ref{com:open})}
The openness of the project can be understood in two ways. Firstly, it is the
openness to new content that is enabling visitors to provide their feedback and
insights to the site by the means of forums, comments etc.

Secondly, once one notices that the user's freedom is restricted by the existing
elements, the issue of the openness to new functionalities arises. If the proper
engine has been chosen it does not seem to be the problem to extend the site by
adding new functionalities to it.

One should remember, however, that the openness does not mean that the site is
going to be extended in unpredicted directions unless they are related to the
site's main subject. This fact distinguishes ''design-on-demand'' sites from
typical Web 2.0 projects where the content and the aim seem to evolve solely
according to users' expectations with no regard given to the goal of site's
creators.

The openness aspect generates one serious threat for the designers. Giving
everyone the possibility to comment and discuss the site's content inevitably
leads to some critical remarks. Some of them are valid and justified and thus
provide valuable feedback. Others, however, are non-constructive and abusive.
They pose a pitfall for the developers as their sole function is to discourage
and intimidate them. The ability to ignore them proves to be critical in site's
continuous evolution and growth.

At the same time the positive feedback from the community is one of the
important motivating factors for the developers. Thus, it is crucial to
encourage users to participate in the project development. This should be
important for project developers to provide some technical means for collecting
community feedback.

\section{Process of building a site}\label{sec:build}
It is important to think of and describe the process of building a
''design-on-demand'' website as it is completely different from the construction
of a typical Web 1.0 or Web 2.0 site.

The process of building a ''design-on-demand'' site is a continuous
effort, which means that it requires prolonged involvement of the developers.
Let us compare this process to the construction of Web 1.0 and Web 2.0 sites
respectively.

In the case of a typical Web 1.0 project, the designers create a site as a final
and complete product. They know from the beginning what the content of
the site will be and once they implement it the process is finished.

In a standard Web 2.0 site case the designers create the framework which, they
hope, would be loaded with content by the contributors/users. They do not have a
vision of a completed site, since the
content is to be provided later by external users. However, these users are
strictly limited by the possibilities which are built into the general framework
of the site.

The ''design-on-demand'' website is created from scratch by its developers, who,
in the beginning, have some general idea about its shape. In time this shape is
modified, extended and amended by the community.

\section{Two examples}\label{sec:examples}
In this section we discuss two examples of web-sites developed using the
''design-on-demand'' methodology.

\subsection{International scientific community}
As the second case we present a Quantiki web portal for scientist
\cite{quantiki}, collecting content relevant for researchers working in quantum
information science.

\paragraph{Subject} In this case the subject was clearly defined from the
beginning and the project was aimed at creating the resource for the quantum
information science community. In the initial phase, however, the project was
developed as a public wiki developed using MediaWiki software and it was aimed
at sharing tutorials, as well as advanced texs related to the subject. During
the lifespan, the goal of the project was redefined and at the moment the
project aim is to aggregate information about activities of the community and to
promote recent research.

\paragraph{People} The initial team of developers consisted of four people,
personally interested in the main topic of the site. At the moment the core team
consists of three people and this proved to be a sufficient number for
maintaining technical, as well as editorial tasks related to the site.

Content is contributed by the relatively large group of users and in this case
it is hard to distinguish between the main and the secondary users. This has one
main drawback. Namely, as many users contribute small portions of content and
provide feedback, it is hard to create hight quality content, which requires
considerable focus.

\paragraph{Promotion} Methods of promoting the discussed portal were specific
for the target community. The portal was mainly advertised during the
international meetings devoted to the main subject. This word-of-mouth promotion
was possible due to the sharply defined target community. Once the community
became aware of the project, many users provided additional support by adding
backlinks to Quantiki on their websites.

\paragraph{Technological background} In the first phase Quantiki was based on
the MediaWiki software which is,at the moment, among the most popular wiki
engines~\cite{wikimatrix}. Unfortunately, the limitations of the software soon
started to interfere with the features requested by the community. This
motivated the migration of some parts of the project to the Drupal content
management system. It should be pointed out that MediaWiki proved to be a good
solution for the purpose of sharing textual based information. However, it does
not provide the mechanisms for seamless sharing and managing more complex data,
\eg\ information about research groups, short video lectures. In the case of
Quantiki the introduction of content management system was directly stimulated
by the needs of the target community. This situation illustrates the importance
of careful planning of the site, since the decisions made at this stage affect
the further development of the project.

\paragraph{Long-term involvement} The project started in the form of public wiki
in May 2005. The target groups of the project were clearly defined from the
beginning. However, the project goals where redefined and, at the moment, its
aim is to collect information about events, research groups and recent progress
in the field of quantum information. Thus the content published on the site it
much more dynamic than in the first phase. This change in the project's aims
occurred after over two years of wiki development.

\paragraph{Openness} The project's main goal was to provide communication
channel for the target-community. Since the initial version, the project was
based on the contributions from the community. Also, as it was already
mentioned, at the moment the majority of contributions are provided by the
community.

\subsection{Vertical portal for travellers}
The first case that illustrates the ''design-on-demand'' approach to creating
web content is Turcja w Sandalach (Turkey in Sandals) portal designed for
independent travellers~\cite{tws}.

\paragraph{Subject} The selection of the subject was a gradual process: from the
initial idea of sharing experience and knowledge with other travellers by
creating a website to the final shape of a portal that not only presents an
individual point of view, but also enables the members of the travel community
to share their stories and exchange recent news. The idea started to evolve in
this direction before any content was added as a result of endless discussions,
brainstorming sessions and research conducted in the web. One of the aims was to
avoid being classified as the n-th 'My holidays in Turkey' story.

\paragraph{People} The basic team of website developers consisted of two people
who were knowledgeable in the main subject of the site. One of them, however,
has had more expertise in technical issues \ie\ CMS systems, registration and
hosting services, while the second one has been focussed on marketing and
promotional activities. This combination has proven to be close-to-perfect, as
the site has been developed seamlessly and continuously, at the same time
avoiding the conflicts arising from the natural differences of opinions.

The supporters followed up, providing essential backup and encouragement as well
as some content. Their engagement proved to be a key motivating factor for the
developers. It gave them the feeling of being noticed and acknowledged by a
community of internet users. It is an astonishing fact that a unique event such
as a single comment or email can provide moral support and a reason for being in
such cases.

Following the initial interest, the traffic statistics were another motivating
element. While a group of supporters is, for obvious reasons, relatively small
yet significant, the number of visitors gives the website developers statistical
motivation \ie\ the sheer number of people visiting the site makes one realise
that the project has gone beyond the initial assumptions and is starting to live
an independent live in the web.

\paragraph{Promotion} The main aim was, from the beginning, to promote the site
at zero-cost level. Access-free forums and catalogues were used as the main
promotion channels. At the first step the designers started advertising the site
by taking part in discussions on the subject that appeared on open forums, at
the same time using the site's address as the signature. This method is usually
available for website promotion as long as the site is not a commercial one.
Otherwise there are some negative consequences, as not only the authors risk
being banned from the given forum, but also lose their credibility as experts in
the area. The second step, that happened practically simultaneously to the first
one, was to register the site in popular directories and catalogues. The goal
here was to increase the site's PageRank and the visibility in the net.

\paragraph{Technological background} The selection of CMS was, in this case,
based on previous experiences of the designers. Additionally, as a non-profit
project, open-source solution was strongly preferred. As the result Drupal CMS
was selected and it has proven to be a flexible, fully expandable and reliable
system.

\paragraph{Long-term involvement} The discussed site has been present in the net
since January 2008. The initial idea was to build a small-sized travel portal,
with limited content and third-part involvement possibilities. As the project
developed, however, it became obvious that its target group had higher
expectations, and the designers responded to this need by extending the site and
offering the users wider opportunities of participation and community-building
activities. The main assumption has always been to add some new content every
week, to make the site look up-to-date.

\paragraph{Openness} As it has been mentioned above the openness for the
travellers' community was, on one hand expressed in the selection of a given CMS
software,and, on the other hand, the community interest in the site has
surpassed the designers' expectations. This case demonstrates how important the
careful selection of technological platform is, as it was the chosen platform
that enabled the designers to extend and meet the community expectations.

\section{Summary}\label{sec:final}
We have presented web community development methodology which is addressed at
medium-sized groups sharing common subject of interests. 

The presented concepts are, in many cases, associated with Web 2.0
term~\cite{bell09building,web20}. However, our aim was to show that, in contrast
to web-projects based on user-generated content, ''design-on-demand'' projects
are shaped by the community. The typical size of the target community in the
discussed type of projects allows for seamless communication between developers
and the community. This distinguishes ''design-on-demand'' project from a
typical Web 2.0 project, as it allows the community to actually influence the
development of the project.

The ability to influence the shape of a project also allows for better community
involvement. It is known that this influences directly the quality of delivered
content~\cite{adamic10individual,nam09questions}. On the other hand, the
''design-on-demand'' project requires initial input from the developers,
specializing in the project's subject. During the project life-time the
developers are responsible for editing content. Thus, the ''design-on-demand''
projects tend to provide higher quality content than in the case of Web 2.0
sites.

Finally the similarity between ''design-on-demand'' methodology and the type of
developer-user relations in the free and open-source community can be
observed~\cite{raymond}. The communication between free software developers and
users is crucial for this type of projects as it enables the developers to
modify software according to users' needs. One the other hand, the positive
feedback provides great motivation for the developers. To some degree, the
''design-on-demand'' method aims to apply a similar methodology to the web
environment.

\bibliographystyle{splncs}
\bibliography{design_on_demand}

\end{document}